\documentstyle[12pt,aasms4]{article}

\begin{document}

\title{POPULATION OF THE SCATTERED KUIPER BELT\footnotemark}
\author{Chadwick A. Trujillo}
\affil{Institute for Astronomy, 2680 Woodlawn Drive, Honolulu, HI
96822 \\ chad@ifa.hawaii.edu}
\author{David C. Jewitt}
\affil{Institute for Astronomy, 2680 Woodlawn Drive, Honolulu, HI
96822 \\ jewitt@ifa.hawaii.edu}
\and
\author{Jane X. Luu}
\affil{Leiden Observatory, PO Box 9513, 2300 RA Leiden, The
Netherlands \\ luu@strw.leidenuniv.nl}
\footnotetext{Based on observations collected at
Canada-France-Hawaii Telescope, which is operated by the National
Research Council of Canada, the Centre National de la Recherche Scientifique
de France, and the University of Hawaii.}

\begin{abstract}
We present the discovery of three new Scattered Kuiper Belt Objects
(SKBOs) from a wide-field survey of the ecliptic.  This continuing
survey has to date covered 20.2 square degrees to a limiting red
magnitude of 23.6.  We combine the data from this new survey with an
existing survey conducted at the University of Hawaii 2.2m telescope
to constrain the number and mass of the SKBOs.  The SKBOs are
characterized by large eccentricities, perihelia near 35 AU, and
semi-major axes $> 50$ AU.  Using a maximum-likelihood model, we
estimate the total number of SKBOs larger than 100 km in diameter to
be $N=(3.1^{+1.9}_{-1.3}) \times 10^{4}$ ($1\sigma$) and the total
mass of SKBOs to be $M \sim 0.05 M_{\oplus}$, demonstrating that the
SKBOs are similar in number and mass to the Kuiper Belt inside 50 AU.
\end{abstract}

\keywords{comets: general --- Kuiper Belt, Oort Cloud --- solar
system: formation}

\section{Introduction}

The outer solar system is richly populated by small bodies in a thick
trans-Neptunian ring known as the Kuiper Belt (Jewitt et al. 1996).
It is widely believed that the Kuiper Belt Objects (KBOs) are remnant
planetesimals from the formation era of the solar system and are
composed of the oldest, least-modified materials in our solar system
(Jewitt 1999).  KBOs may also supply the short-period comets
(Fern\'{a}ndez and Ip 1983; and Duncan et al. 1988).  The orbits of
KBOs are not randomly distributed within the Belt but can be grouped
into three classes.  The Classical KBOs constitute about 2/3 of the
known objects and display modest eccentricities ($e \sim 0.1$) and
semi-major axes ($41 \mbox{ AU} < a < 46 \mbox{ AU}$) that guarantee a
large separation from Neptune at all times.  The Resonant KBOs, which
comprise most of the remaining known objects, are trapped in
mean-motion resonances with Neptune, principally the 3:2 resonance at
39.4 AU.  In 1996, we discovered the first example of a third
dynamical class, the Scattered KBOs (SKBOs): $\rm 1996 TL_{66}$
occupies a large ($a \sim 85$ AU), highly eccentric ($e \sim 0.6$),
and inclined ($i \sim 25^{\circ}$) orbit (Luu et al. 1997).  It is the
defining member of the SKBOs, characterized by large eccentricities
and perihelia near $35$ AU (Duncan and Levison 1997).  The SKBOs are
thought to originate from Neptune scattering, as evidenced by the fact
that all SKBOs found to date have perihelia in a small range near
Neptune ($34 \mbox{ AU} < q < 36 \mbox{ AU}$).

In this letter we report preliminary results from a new survey of the
Kuiper Belt undertaken at the Canada-France-Hawaii Telescope (CFHT)
using a large format Charge-Coupled Device (CCD).  This survey has
yielded 3 new SKBOs.  Combined with published results from an earlier
survey (Jewitt et al. 1998) and with dynamical models (Duncan and
Levison 1997), we obtain new estimates of the population statistics of
the SKBOs.  The main classes of KBOs appear in Figure~\ref{plan}, a
plan view of the outer solar system.

\section{Survey Data}

Observations were made with the 3.6m CFHT and 12288 x 8192 pixel
Mosaic CCD (Cuillandre et al., in preparation).  Ecliptic fields were
imaged within 1.5 hours of opposition to enhance the apparent
sky-plane speed difference between the distant KBOs ($\sim 3$ arc
sec/hr) and foreground main-belt asteroids ($\gtrsim 25$ arc sec/hr).
Parameters of the CFHT survey are summarized in Table~\ref{obs}, where
they are compared with parameters of the survey earlier used to
identify $\rm 1996 TL_{66}$ (Luu et al. 1997 and Jewitt et al. 1998).
We include both surveys in our analysis of the SKBO population.

Artificial moving objects were added to the data to quantify the
sensitivity of the moving object detection procedure (Trujillo and
Jewitt 1998).  The seeing in the survey varied from 0.7 arc sec to 1.1
arc sec (FWHM).  Accordingly, we analysed the data in 3 groups based
on the seeing.  Artificial moving objects were added to
bias-subtracted twilight sky-flattened images, with profiles matched
to the characteristic point-spread function for each image group.
These images were then passed through the data analysis pipeline.  The
detection efficiency was found to be uniform with respect to sky-plane
speed in the 1 -- 10 arc sec/hr range, with efficiency variations due
only to object brightness.

The seeing-corrected efficiency function was fitted with a hyperbolic
tangent profile for use in the maximum-likelihood SKBO orbital
simulation described in Section~\ref{ml-section}:
\begin{equation}
\varepsilon =  \frac{\varepsilon_{\rm max}}{2} \left( \tanh \left( \frac{m_{\rm R50} -
m_{\rm R}}{\sigma} \right) + 1 \right) ,
\end{equation}
where $0 < \varepsilon < 1$ is the efficiency at detecting objects of
red magnitude $m_{\rm R}$, $\varepsilon_{\rm max} = 0.83$ is the
maximum efficiency, $m_{\rm R50} = 23.6$ is the magnitude at which
$\varepsilon = \varepsilon_{\rm max}/2$, and $\sigma = 0.4$ magnitudes
is the characteristic range over which the efficiency drops from
$\varepsilon_{\rm max}$ to zero.

The orbital parameters of the 4 established SKBOs are listed in
Table~\ref{skbos}.  It should be noted that the 3 most recent KBOs
have been observed for a timebase of about 3 months, so their fitted
orbits are subject to revision.  However, the 4 year orbit of $\rm
1996 TL_{66}$ has changed very little since its initial announcement
with a 3 month arc ($a = 85.754$, $e = 0.594$, $i = 23.9$, $\omega =
187.7$, $\Omega = 217.8$, $M = 357.3$, and epoch 1996 Nov 13; Marsden
1997).  A plot of eccentricity versus semi-major axis appears in
Figure~\ref{evsa}, showing the dynamical distinction of SKBOs from the
other KBOs.  All SKBOs have perihelia $34 \mbox{ AU} < q < 36 \mbox{
AU}$.  To avoid confusion with Classical and Resonant KBOs also having
perihelia in this range, we concentrate on the SKBOs with semi-major
axes $a > 50$ AU.  In so doing, we guarantee that our estimates
provide a solid lower bound to the true population of SKBOs.

\section{Population Estimates}
\label{ml-section}

The number of SKBOs can be crudely estimated by simple extrapolation
from the discovered objects.  The faintest SKBO in our sample, $\rm
1999 CY_{118}$, was bright enough for detection only during the 0.24\%
of its $\sim 1000$ year orbit spent near perihelion.  Our 20.2 square
degrees of ecliptic observations represent $\sim 1/500$th of the total
$\pm 15^{\circ}$ thick ecliptic, roughly the thickness of the Kuiper
Belt (Jewitt et al. 1996).  Thus, we crudely infer the population of
SKBOs to be of order $500 / 2.4 \times 10^{-3} \sim 2 \times 10^{5}$.
To more accurately quantify the population, we use a
maximum-likelihood method to simulate the observational biases of our
survey.

Our discovery of 4 SKBOs with the CFHT and UH telescopes provides the
data for the maximum-likelihood simulation.  We assess the intrinsic
population of SKBOs by the following procedure (c.f. Trujillo 2000):
(1) create a large number ($\sim 10^{8}$) of artificial SKBO orbits
drawn from an assumed distribution; (2) determine the ecliptic
latitude, longitude, and sky-plane velocity of each artificial object,
using the equations of Sykes and Moynihan (1996; a sign error was
found in equation 2 of their text and corrected before use), and
compute the object's brightness using the $H$, $G$ formalism of Bowell
et al. (1989), assuming an albedo of 0.04; (3) determine the
detectability of each artificial object based on the detection
efficiency and sky-area imaged in the two surveys; (4) create a
histogram of ``detected'' artificial objects in $a$-$e$-$i$-radius
space, with a bin size sufficiently small such that binning effects
are negligible; (5) based on this histogram, compute the probability
of finding the true, binned, observed distribution of the four SKBOs,
assuming Poisson detection statistics; and (6) repeat the preceding
steps, varying the total number of objects in the distribution ($N$)
and the slope $(q' = 3, 4)$ of the differential size distribution to
find the model most likely to produce the observations.  A full
description of the model parameters is found in Table~\ref{model}.

This model was designed to match the SKBO population inclination
distribution, eccentricity distribution, and ecliptic plane surface
density as a function of heliocentric distance, as found in the only
published SKBO dynamical simulation to date, that of Duncan and
Levison (1997).  It should be noted that the results of the 4 billion
year simulations of Duncan and Levison (1997) were based on only 20
surviving particles, averaged over the last 1 billion years of their
simulation for improved statistical accuracy.  The total number of
objects and the size distribution of objects remained free parameters
in our model.

\section{Results}

The results of the maximum-likelihood simulation appear in
Figure~\ref{ml}.  Confidence levels were placed by finding the
interval over which the integrated probability distribution
corresponded to 68.27\% and 99.73\% of the total, hereafter referred
to as $1\sigma$ and $3\sigma$ limits, respectively.  The total number
of SKBOs with radii between 50 km and 1000 km is
$N=(3.1^{+1.9}_{-1.3}) \times 10^{4}$ ($1\sigma$) in the $q' = 4$
case, with $4.0 \times 10^{3} < N < 1.1 \times 10^{5}$ $3\sigma$
limits.  The $q' = 3$ case is similar with $N = (1.4^{+1.1}_{-0.5})
\times 10^{4}$ ($1\sigma$), with $2.0 \times 10^{3} < N < 5.3 \times
10^{4}$ $3\sigma$ limits.  The $q'= 4$ and $q'= 3$ best fits are
equally probable at the $<1\sigma$ level, however we prefer the $q'=
4$ case as recent measurements of the Kuiper Belt support this value
(Jewitt et al. 1998, Gladman and Kavelaars 1998, Chiang and Brown,
1999).  This population is similar in number to the Kuiper Belt
interior to 50 AU, which contains about $\sim 10^{5}$ objects (Jewitt
et al. 1998).  The observation that only a few percent of KBOs are
SKBOs is due to the fact that the SKBOs are only visible in
magnitude-limited surveys during the small fraction of their orbits
when near perihelion.  In addition, the high perihelion velocities,
$v$, of eccentric SKBOs may have implications for erosion in the
Kuiper Belt since ejecta mass, $m_{\rm e}$, scales with impact energy,
$E$, as $m_{\rm e} \sim E^{2} \sim v^{4}$ (Fujiwara et al. 1977).

Extrapolating the $q'= 4$ distribution to the size range $1 \mbox{ km}
< r < 10 \mbox{ km}$ yields $N \sim 4 \times 10^9$ SKBOs in the same
size range as the observed cometary nuclei.  This is comparable to the
$10^{9}$ needed for the SKBOs to act as the sole source of the short-period comets (Levison and Duncan 1997).

The total mass $M$ of objects assuming $q'= 4$ is
\begin{equation}
M = 2.3 \times 10^{-3} p_{R}^{3/2} \frac{\rho}{\mbox{kg }\mbox{m}^{-3}} \log(\frac{r_{\rm max}}{r_{\rm min}}) M_{\oplus},
\end{equation}
where red geometric albedo $p_{\rm R} = 0.04$, density $\rho = 2000
\mbox{ kg m}^{-3}$, the largest object $r_{\rm max} = 1000$ km
(Pluto-sized), and smallest object $r_{\rm min} = 50$ km yields $M
\sim 0.05 M_{\oplus}$, where $M_{\oplus} = 6 \times 10^{24}$ kg is the
mass of the earth.  This is comparable to the total mass of the
Classical and Resonant KBOs (0.2 $M_{\oplus}$, Jewitt et al. 1998).
If only 1\% percent of SKBOs have survived for the age of the solar
system (Duncan and Levison 1997), then an initial population of $10^7$
SKBOs ($50 \mbox{ km} < r < 1000 \mbox{ km}$) and a SKBO primordial
mass $\gtrsim 5 M_{\oplus}$ are inferred.

As our observations are concentrated near the ecliptic, we expected
most of the discovered objects to have low inclinations.  However, of
the 4 established SKBOs, 3 have high inclinations ($i \gtrsim
20^{\circ}$).  The possibility of a Gaussian distribution of
inclinations centered at $i \sim 20^{\circ}$ was tested, however, the
fit was not better than the uniform $i$ case at the $\geq\!3\sigma$
level.  If the SKBOs were distributed as such, the total numbers of
SKBOs reported here would be enhanced by a factor $\sim 2$.  Combining
this model-dependent uncertainty, the orbital uncertainties of the 4
SKBOs, and the computed Poisson noise, we estimate that our
predictions provide an order-of-magnitude constraint on the population
of the SKBOs.  As additional SKBOs become evident in our continuing
survey and as recovery observations of the SKBOs are secured, we will
refine this estimate.

\section{Summary}

We have constrained the SKBO population using surveys undertaken on
Mauna Kea with two CCD mosaic cameras on two telescopes.  The surveys
cover 20.2 square degrees to limiting red magnitude 23.6 and 51.5
square degrees to limiting red magnitude 22.5.  We find the following:

(1) The SKBOs exist in numbers comparable to the KBOs.  The
    observations are consistent with a population of
    $N=(3.1^{+1.9}_{-1.3}) \times 10^{4}$ ($1\sigma$) SKBOs in the
    radius range $50 \mbox{ km} < r < 1000 \mbox { km}$ with a
    differential power-law size distribution exponent of $q' = 4$.

(2) The SKBO population in the size range of cometary nuclei is large
    enough to be the source of the short-period comets.

(3) The present mass of the SKBOs is approximately $0.05 M_{\oplus}$.
    When corrected for depletion inferred from dynamical models
    (Duncan and Levison 1997), the initial mass of SKBOs in the early
    solar system approached $5 M_{\oplus}$.

We are continuing our searches for these unusual objects as it is clear
that they play an important role in outer solar system dynamics.

\acknowledgments

We thank Ken Barton for help at the Canada-France-Hawaii Telescope.
This work was supported by a grant to DCJ from NASA.

\newpage

\section{Figure Captions}

\figcaption{A plan view of the outer solar system; the green circle
represents the orbit of Jupiter.  Kuiper Belt Objects have been
color-coded based on their orbital classification.  The SKBO orbits
are distinct. \label{plan}}


\figcaption{The KBOs (hollow) and the 4 SKBOs (solid).  The 3:2 and
2:1 resonances are shown as dashed lines.  The solid lines represent
perihelia of $q$ = 34 and 36. \label{evsa}}

\figcaption{Probability that the model matches the observed distribution
of SKBOs versus the total number of objects in the model for the $q'$
= 4 size distribution.  The maximum likelihood is given by
$N=(3.1^{+1.9}_{-1.3}) \times 10^{4}$ ($1\sigma$).  The dashed lines
represent the $1 \sigma$ and $3 \sigma$ limits. \label{ml}}

\begin{center}
\begin{deluxetable}{ccc}
\tablecaption{Survey Parameters}
\tablehead{\colhead{Quantity} & \colhead{UH 2.2m} & \colhead{CFHT 3.6m}}
\startdata
Focal Ratio & f/10 & f/4 \nl
Instrument & UH 8k Mosaic & CFHT 12k x 8k Mosaic \nl
Plate Scale [arc sec/pixel] & 0.135 & 0.206 \nl
Field Area [$\rm deg^{2}$] & 0.09 & 0.33 \nl
Total Area [$\rm deg^{2}$] & 51.5 & 20.2\nl
$m_{\rm R50}$\tablenotemark{a} & 22.5 & 23.6 \nl
$\theta$\tablenotemark{b} [arc sec] & 0.8--1.0 & 0.7--1.1 \nl
Filter & $V\!R\,$\tablenotemark{c} & $R$ \nl
Quantum Efficiency & 0.33 & 0.75 \nl
$N_{\rm SKBOs}$\tablenotemark{d} & 1 & 3 \nl
\enddata
\tablenotetext{a}{The red magnitude at which detection efficiency reaches half of the maximum efficiency.}
\tablenotetext{b}{the typical Full Width at Half Maximum of stellar sources
for the surveys.}
\tablenotetext{c}{The $V\!R$ filter has a high transmission in the 5000
-- 7000 $\AA$ range (Figure 1 of Jewitt et al., 1996).}
\tablenotetext{d}{The number of SKBOs detected.}
\label{obs}
\end{deluxetable}
\end{center}

\begin{center}
\begin{deluxetable}{ccccc}
\tablecaption{The SKBOs\tablenotemark{*}}
\tablehead{ \colhead{} & \colhead{$\rm 1996 TL_{66}$} & \colhead{$\rm 1999 CV_{118}$} & \colhead{$\rm 1999 CY_{118}$} & \colhead{$\rm 1999 CF_{119}$}  }
\startdata					      				                                                                      
Orbital Properties \nl				      				                                                                      
$a$ [AU]               & 85.369                       & 56.532                        & 95.282                        & 115                            \nl
$e$                    & 0.590                        & 0.387                         & 0.642                         & 0.687                          \nl
$i$ [deg]              & 23.9                         & 5.5                           & 25.6                          & 19.7                           \nl
$\omega$ [deg]         & 184.5                        & 126.6                         & 21.4                          & 218.1                          \nl
$\Omega$ [deg]         & 217.8                        & 305.6                         & 163.1                         & 303.4                          \nl
$M$ [deg]              & 359.2                        & 22.5                          & 356.4                         & 354.3                          \nl
epoch                  & 1999 Aug 10                  & 1999 Mar 23                   & 1999 Mar 23                   & 1999 Mar 23                    \nl
\nl						      				                                                                      
Discovery Conditions \nl			      				                                                                      
date                   & 1996 Oct 09                  & 1999 Feb 10                   & 1999 Feb 10                   & 1999 Feb 11                    \nl
$R$ [AU]               & 35.2                         & 39.7                          & 38.6                          & 42.2                           \nl
$m_{\rm R}$            & 20.9                         & 23.0                          & 23.7                          & 23.0                           \nl
$m_{\rm R}(1,1,0)$     & 5.3                          & 7.0                           & 7.8                           & 6.7                            \nl
Diameter [km]          & 520                          & 240                           & 170                           & 270                            \nl
\enddata

\tablecomments{$a$, $e$, $i$, $\omega$, $\Omega$, and $M$ represent
the Keplerian orbital elements semi-major axis, eccentricity,
inclination, argument of perihelion, longitude of the ascending node,
and mean anomaly at the given epoch.  $R$ is the heliocentric
distance. $m_{\rm R}$ is the red magnitude at discovery.  $m_{\rm
R}(1,1,0)$ is red magnitude at zero phase angle, geocentric distance =
$R = 1$ AU, assuming a phase darkening law for dark objects (Bowell,
et al. 1989).  The diameter is derived assuming a geometric albedo of
4\%.}

\tablenotetext{*}{Orbital elements from B. Marsden, Minor Planet
Center, Harvard-Smithsonian Center for Astrophysics.}
\label{skbos}
\end{deluxetable}
\end{center}

\begin{center}
\begin{deluxetable}{cccl}
\tablecaption{Model Parameters}
\tablehead{\colhead{Symbol} & \colhead{Value} & \colhead{Distribution} & \colhead{Description}}
\startdata
$a$                         & 50 -- 200 AU    & note\tablenotemark{a} & semi-major axis\nl
$e$                         & 0 -- 1          & uniform & eccentricity \nl
$i$                         & 0 -- 35 deg     & uniform & inclination\nl
$\omega$                    & 0 -- 360 deg    & uniform & argument of perihelion\nl
$\Omega$                    & 0 -- 360 deg    & uniform & longitude of the ascending node\nl
$M$                         & 0 -- 360 deg    & uniform & mean anomaly \nl
$r$                         & 50 -- 1000 km   & $n(r) dr \sim r^{-q'} dr$& radius\nl
$q'$                        & 3, 4            & --- & slope parameter \nl
$p_{\rm R}$                 & 0.04            & --- & red albedo \nl
$N$                         & free            & --- & total number of objects\nl
$q$                         & 34 -- 36 AU     & note\tablenotemark{b} & perihelion distance \nl
\enddata
\tablenotetext{a}{Semi-major axis was chosen such that the surface
density of objects versus heliocentric distance follows $n(R) \sim
R^{-p} dR$, where $p = 2.5$ (Duncan and Levison 1997).}
\tablenotetext{b}{Perihelion distance was selected by first choosing
$a$ and then limiting $e$ such that objects were in the 34 to 36 AU range.}
\label{model}
\end{deluxetable}
\end{center}

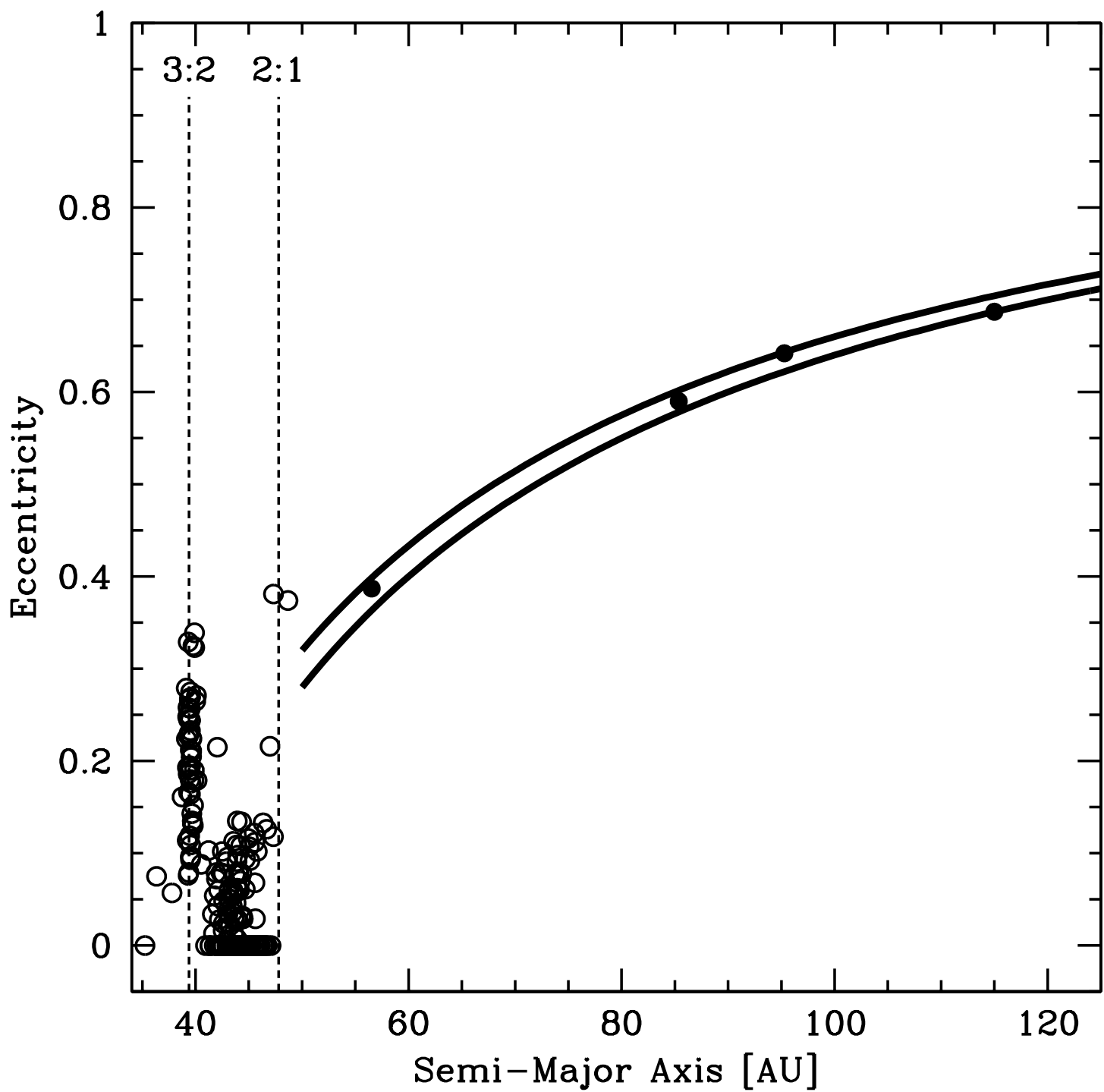
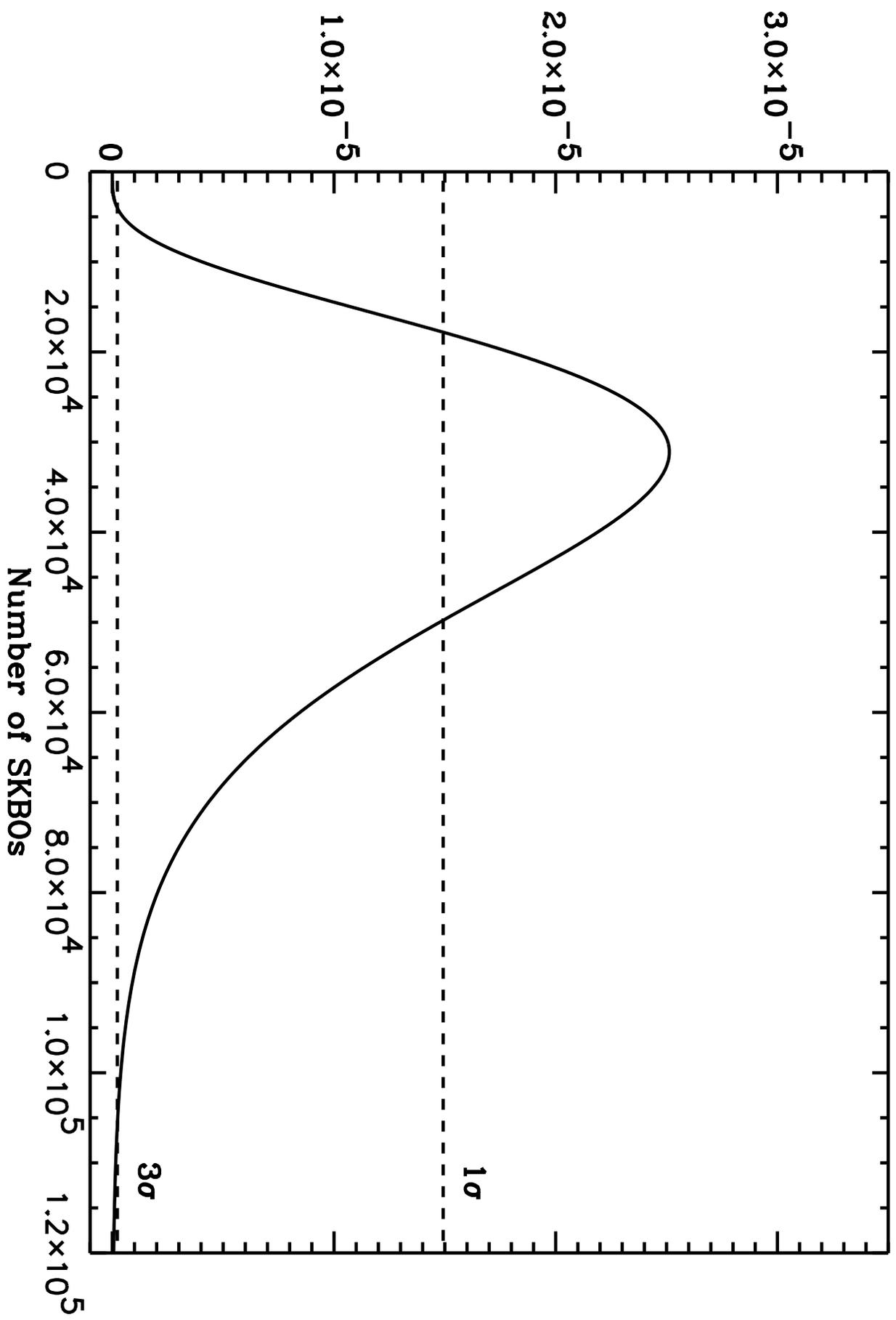
\end{document}